\documentstyle[aps,prl,twocolumn]{revtex}

\begin{document}

\title{A toy model for molecular condensates in Bose gases}

\author{L.~Pricoupenko}

\address{Laboratoire de Physique Th{\'e}orique des Liquides
Universit{\'e} Pierre et Marie Curie, 4 place Jussieu, 75252 Paris Cedex
05, France.
\thanks{L.P.T.L. is UMR n$^{\circ}$7600 associ{\'e}e au CNRS.}}

\date{\today}

\maketitle

\begin{abstract}
The occurrence of a molecular Bose-Einstein condensate is studied for
an atomic system near a zero energy resonance of the binary
scattering process, with a large and positive scattering length. 
The interaction potential is modeled by a pseudo-potential having one bound state.
Using a variational Gaussian ansatz for the $N$-body density operator, 
we discuss the thermodynamic properties at low temperature and the relative stability
of the system towards the formation of an atomic Bose-Einstein
condensate. We also derive an approximate Gross-Pitaevskii equation for the
molecular condensate leading to the prediction of a Bogoliubov spectrum.
\end{abstract}

The atomic Bose-Einstein condensation has been extensively
studied during past years both theoretically and
experimentally~\cite{BEC,Revue}. In this Letter, we consider the
different case of a molecular Bose-Einstein condensate. 
The realization of such a situation is a challenge in the
field of low temperature physics. Already cold gases of molecules have
been produced \cite{Boston,Pillet,Heinzen}. No evidence for a molecular
condensate has been obtained yet, although the stimulated Raman technique
used in \cite{Heinzen}, forming molecules directly from an atomic
condensate, is promising. 

	Another scenario is possible for the formation of the molecular
condensate. Consider the case of a so-called {\sl zero-energy} resonance
in the two-body scattering process, where the {\bf positive} scattering
length $a>0$ is much larger than the effective range $r_e$ of the interaction
potential \cite{Landau}. In this case the two-body potential supports
a $s$-wave bound state  $\phi_0$ with a spatial extension $\simeq a$ much
larger than the other $s$-wave bound states and with a much smaller
binding energy $E_0=-\hbar^2/(ma^2)$. One then would rely simply on
three-body collisions between atoms of a trapped atomic condensate
to produce molecules in the state $\phi_0$. As shown in \cite{Shlyap,Greene}
the rate of formation of molecules scales as $a^4$ in the limit of large $a$
and mainly leads to formation of molecules in the highest $s$-wave two-body
channel \cite{Greene} that is in the state $\phi_0$. From energy conservation
one finds that the molecules produced in $\phi_0$ have a kinetic energy
$-E_0/3$ when the initial atomic wavevectors $\vec{k}$ are such that $k a\ll 1$.
The requirement that the molecules formed in state $\phi_0$ remain in the
trap imposes a depth of the trapping potential larger than $-E_0$. For
a modest trap depth of 10~$\mu$K achievable in an optical trap
\cite{Ketterle_dipole} and for the mass of Rubidium one finds that $a$
has to be larger than 445 Bohr radii. Such high values of $a$ may be 
obtained using a Feshbach resonance induced by a magnetic field
\cite{Ketterle_Feshbach,Cornell_Feshbach}.

In this Letter we assume that a condensate of molecules in the
diatomic bound state $\phi_0$ has been formed. We describe the molecular
condensate at the atomic level: we take as a starting point a model Hamiltonian
for interacting atoms and we use a Gaussian variational ansatz for the
$N-$body density operator including only the effect of binary interactions
between atoms. In the low density regime ($n a^3\ll 1$ where $n$ is the
density of atoms) we determine (i) the properties of the molecular
condensate at thermal equilibrium and (ii) the response of the
molecular condensate to a small perturbation of the trapping potential.
Our results for both cases correspond formally to the known properties of a weakly
interacting Bose-Einstein condensate of particles of mass $2m$ 
having a coupling constant $6g$, where $m$ is the atomic mass and $g=4\pi\hbar^2a/m$ is the atomic
coupling constant. At higher density ($n a^3>\pi/192$) we find that an atomic condensate forms.
To model the binary atomic interaction potential
we use the pseudo-potential $V$ defined by the following 
action on a two-body atomic wavefunction $\psi$:
\begin{equation}
\langle\vec{r}_1,\vec{r}_2|V|\psi\rangle = 
g \delta(\vec{r}\,) \partial_r\left[r \psi(\vec{R}-\vec{r}/2,\vec{R}+\vec{r}/2)\right] \quad , \label{eq:vpseudo} 
\end{equation}
where we have introduced the coordinates of the center of mass 
and of the relative motion:
\begin{equation}
 \vec{R} = \frac{\vec{r}_1+\vec{r}_2}{2}, \quad 
 \vec{r} = \vec{r}_1 - \vec{r}_2 \quad \text{and} \quad 
r = \| \vec{r}\, \| \quad .
\end{equation}
This model potential discussed in \cite{Huang} 
has been used to extend the BCS theory to
inhomogeneous atomic systems \cite{BCS}. As we now see,
it has the ability to capture the essential feature of a  general
binary scattering problem near a zero-energy resonance.
Consider indeed the relative motion of two atoms 
described by the hamiltonian
\begin{equation}
	{\cal H}_r = -\frac{\hbar^2}{m}\Delta_{\vec{r}} + V \quad .
\end{equation}
Contrary to the case of the usual contact interaction, the
scattering of two atoms interacting with the potential
(\ref{eq:vpseudo}) is a well defined problem. The scattering amplitude
for a relative wavevector $\vec{k}$ is
given by $-a/(1+ika)$ reproducing the known universal Lorentzian
shape of the scattering cross section near a zero energy resonance 
\cite{Landau}. In addition to the usual scattering states, the
pseudo-potential for $a>0$ leads to the asymptotic form of the $s$-wave
bound state of energy $E_0$:
\begin{equation}
\phi_0(\vec{r}\,) = \frac{1}{r\ \left(2\pi a\right)^{1/2}}
\exp\left(-\frac{r}{a}\right) \quad .
\label{eq:bound}
\end{equation} 
This property of the pseudo-potential reproduces the universal
behaviour for a general interaction potential close to a zero-energy 
resonance, already mentioned in the introduction. It plays an
essential role in our approach as it allows the formation of pairs of
atoms, that is binary molecules, while keeping the mathematical
simplicity of a zero range model potential ($r_e=0$).

	The goal of this paper is to study the properties of a
molecular condensate, in a symmetry breaking approach, this
corresponds to $\langle\hat{\Psi}\rangle=0$ and $\langle\hat{\Psi}\hat{\Psi}\rangle \neq 0$, where
$\hat{\Psi}$ is the atomic field operator. However it is dangerous to
exclude {\sl a priori} the coexistence of an atomic and a molecular
condensate. We therefore use the more general symmetry breaking
prescription by splitting the field operator in a classical part 
and a quantum fluctuation part: $\hat{\Psi} = \Phi + \hat{\phi}$, with 
$\langle\hat{\phi}\rangle = 0$. We choose a Gaussian ansatz for the many-body 
density matrix  $D = Z^{-1} \exp - \beta K $ with 
\begin{eqnarray}
K &=&  \int\!d^3r_1 d^3r_2 \hat{\phi}^{\dagger}(\vec{r}_1)
h(\vec{r}_1,\vec{r}_2) \hat{\phi}(\vec{r}_2) \nonumber\\
& & + \frac{1}{2}\int\!d^3r_1 d^3r_2  [\hat{\phi}(\vec{r}_1)
\Delta(\vec{r}_1,\vec{r}_2) \hat{\phi}(\vec{r}_2) + {\rm h.c.}] \quad .\nonumber
\end{eqnarray}
 	There are three variational fields in this theory: $h,\Delta$ and
$\Phi$. At thermal equilibrium, the grand potential $E-\mu N - TS$ is minimum
so that the functions $h(\vec{r}_1,\vec{r}_2)$ and
$\Delta(\vec{r}_1,\vec{r}_2)$ may be expressed in terms of the one-body density matrix  
$\bar \rho(\vec{r}_1,\vec{r}_2)=\langle \hat{\phi}^\dagger(\vec{r}_2)
\hat{\phi}(\vec{r}_1)\rangle$ and of the pairing function $\bar \kappa(\vec{r}_1,\vec{r}_2)=\langle \hat{\phi}(\vec{r}_1)
\hat{\phi}(\vec{r}_2)\rangle$, while $\Phi(\vec{r}\,)$ verifies a partial differential equation\cite{Blaizot}
\begin{eqnarray}
h(\vec{r}_1,\vec{r}_2) &=& -\frac{\hbar^2}{2m}(\vec{\nabla}^2\delta)(\vec{r}\,)
+ (2g n(\vec{R}\,) - \mu )\delta(\vec{r}\,) \label{eq:h_eq}\\
\Delta(\vec{r}_1,\vec{r}_2) &=& g \delta(\vec{r}\,) [\bar \kappa_{\rm reg}(\vec{R}\,)
+ \Phi^2(\vec{R}\,)]
\label{eq:delta_eq}\\
-\frac{\hbar^2}{2m} \Delta \Phi &+& [g(2n-|\Phi|^2)-\mu] \Phi + g \bar \kappa_{\rm reg} \Phi^*
= 0 \label{eq:Phi}
\end{eqnarray}
In these equations, we  have introduced  the atomic density at a point
$\vec{R}$ 
\begin{equation}
n(\vec{R}\,)=  |\Phi|^2(\vec{R}\,) + \bar \rho(\vec{R},\vec{R}\,) \label{eq:densat}
\quad ,
\end{equation}
and the regular part of the pairing function 
\begin{equation}
\bar \kappa_{\rm reg}(\vec{R}\,) = \lim_{r\to 0}
\partial_r\left[r \bar \kappa(\vec{R}-\vec{r}/2,\vec{R}+\vec{r}/2)\right] \quad .
\label{eq:kappa_reg}
\end{equation}
We assume here that the atoms are in a cubic box of size $L$
with periodic boundary conditions \cite{Trap}, so that 
$\Phi, n, \kappa_{\rm reg}$ do not depend on position. We expand the field
operator on plane waves using a Bogoliubov transform:
\begin{equation}
\hat{\phi}(\vec{r}\,) = \frac{1}{L^{3/2}} \sum_{\vec{k}} \hat{b}_{\vec{k}} u_k \exp ( i
\vec{k}\cdot \vec{r}\,) + \hat{b}_{\vec{k}}^\dagger v_k^* \exp (- i \vec{k}\cdot
\vec{r}\,) .
\end{equation}
The commutation relations of the bosonic annihilation operators
$\hat{b}_{\vec{k}}$'s lead to the normalization of the modes amplitudes
$|u_k|^2 - |v_k|^2 = 1$. We search the $(u_k,v_k)$'s so that $K$ is a
sum  of decoupled harmonic oscillators:
\begin{equation}
	K=K_0 + \sum_{\vec{k}} \hbar \omega_k \hat{b}_{\vec{k}}^\dagger\hat{b}_{\vec{k}} \quad .
\label{eq:KBB}
\end{equation}
From the equilibrium conditions (\ref{eq:h_eq},\ref{eq:delta_eq}), 
we find that each $(u_k,v_k)$ is the eigenvector of the system:
\begin{eqnarray}
 h_k u_k +  g (\Phi^2+ \bar \kappa_{\rm reg}) v_k &=&
\hbar\omega_k u_k \nonumber \\
 h_k v_k + g (\Phi^2+\bar \kappa_{\rm reg}) u_k &=& - \hbar\omega_k v_k 
\label{eq:valpropres}
\end{eqnarray}
with the notation $h_k= \displaystyle\frac{\hbar^2 k^2}{2m} + 2gn -
\mu$. A simple algebra gives the expressions for the eigenvectors 
and the spectrum $\{\hbar \omega_k\}$ \cite{warning}:
\begin{eqnarray}
v_k^2 &=& \frac{1}{2} \left[ \frac{h_k}{\hbar\omega_k} - 1 \right] \\
\hbar\omega_k &=& \left[ h_k^2-g^2(\Phi^2+\bar \kappa_{\rm reg})^2 \right]^{1/2}
\label{eq:spectrum}
\end{eqnarray}
The pairing function is given by
\begin{equation}
\bar \kappa(\vec{r}_1,\vec{r}_2) = - \frac{g (\bar \kappa_{\rm
reg}+ \Phi^2)}{2L^3}\sum_{\vec{k}} \frac{1+2f_k}{\hbar \omega_k} \exp(i \vec{k}\cdot\vec{r}\,) \quad ,\label{eq:kappa_eq} \\
\end{equation}
and the one-body density matrix  
\begin{equation}
\bar \rho(\vec{r}_1,\vec{r}_2) = \frac{1}{L^3}\sum_{\vec{k}}\left[\left(2f_k+1\right)
v_k^2+f_k\right] \exp(i \vec{k}\cdot\vec{r}\,) \quad .\label{eq:rho_eq}
\end{equation}
In Eqs.(\ref{eq:kappa_eq},\ref{eq:rho_eq}), $f_k$ is 
the Bose occupation factor $f_k = [\exp ( \beta \hbar \omega_k ) - 1 ]^{-1}$.

	All the equilibrium properties may be expressed in terms of
$(n,T)$. For this purpose, we have to determine the three unknown
parameters $(\mu,\Phi,\bar \kappa_{\rm reg})$. In the presence of an atomic
condensate, $\Phi$ does not vanish and a first relation is given by 
Eq.(\ref{eq:Phi}) 
\begin{equation}
\mu = g [2n+\bar \kappa_{\rm reg} - \Phi^2] \quad .
\label{eq:mu_Phi}
\end{equation}
A second one is obtained from Eq.(\ref{eq:densat}), by
setting $\vec{r}=0$ in Eq.(\ref{eq:rho_eq})
\begin{equation}
 n = \Phi^2 + \frac{1}{L^3}\sum_{\vec{k}}\left[\left(2f_k+1\right)
v_k^2+f_k\right] \quad .\label{eq:norm}
\end{equation}
 The third equation is obtained by extracting the regular part of 
Eq.(\ref{eq:kappa_eq}) as in \cite{BCS}; this leads to
\begin{equation}
\bar \kappa_{\rm reg} = \frac{g(\Phi^2+\bar \kappa_{\rm reg})}{L^3}\sum_{\vec{k}}\left[\frac{m}{\hbar^2
k^2}-\frac{1+2f_k}{2\hbar\omega_k} \right] \label{eq:gap}.
\end{equation}
In what follows, we consider the thermodynamical limit so that the
sums over $\vec{k}$ are replaced by
integrals, we also restrict to the case of a vanishing temperature
$T=0$, so that the Bose occupation factors $f_k$ is zero. 

	First, we note that the stability of the ground state imposes 
real values for the spectrum $\{\hbar \omega_k\}$, this implies $\bar \kappa_{\rm reg} < 0$ and from 
Eq.(\ref{eq:gap}), we find $|\bar  \kappa_{\rm reg}| > |\Phi|^2$. In particular
for a vanishing number of atoms in the condensate $\Phi=0$, we get from
Eqs.(\ref{eq:mu_Phi},\ref{eq:norm},\ref{eq:gap}):
\begin{equation}
	\bar \kappa_{\rm reg}^{\rm c} = - \frac{\pi}{64a^3} \quad , \quad
n^{\rm c} = \frac{\pi}{192a^3} \quad .
\end{equation}
	This value of the density determines the threshold of coexistence of an atomic
condensate ($\Phi \neq 0$) with a molecular one. For $n>n^c$, the model
predicts the coexistence, a result already obtained in a slightly
different approach in \cite{Nozieres}. In this high density regime, our 
modelization is questionable and we do not pursue the study in
this range anymore. 

	For $n<n^{\rm c}$, there is no atomic condensate: $\Phi$ is
identically zero so that Eq.(\ref{eq:mu_Phi}) does not hold in this regime and equilibrium
properties are deduced from Eqs.(\ref{eq:norm},\ref{eq:gap}) only. 
We now wish to check that the low density regime corresponds
indeed to a molecular condensate. For this purpose, we suppose that 
the chemical potential tends to a finite negative value, hence 
$|\mu|\gg gn,g|\bar \kappa_{\rm reg}|$ and $\hbar\omega_k \simeq \hbar^2k^2/(2m)-\mu$.
From Eq.(\ref{eq:gap}) we find the lowest order approximation 
\begin{equation}
\mu \simeq  \frac{E_0}{2} \quad, \label{eq:brisure} 
\end{equation}
which is finite indeed, and negative. This result is enlightening:
$E_0$ is just the binding energy of a pair of atoms in the bound state
Eq.(\ref{eq:bound}) so that the value Eq.(\ref{eq:brisure}) of the chemical potential corresponds
to that of an ideal gas of molecules. Since we are at zero temperature
this gas of pairs is actually a molecular Bose-Einstein condensate.

	This interpretation is confirmed by the lowest order approximation 
to the pairing function $\kappa$ in the low density limit. One first
calculates $\bar \kappa_{\rm reg}$ from Eq.(\ref{eq:norm}) using the lowest order approximation 
$v_k\simeq g \bar \kappa_{\rm reg}(E_0-\hbar^2 k^2/m)^{-1}$ \cite{gap}: 
\begin{equation}
\bar \kappa_{\rm reg} \simeq -\frac{1}{a^3}\left(\frac{na^3}{2\pi}\right)^{1/2}
\quad ,
\end{equation}
then we calculate the integral over $\vec{k}$ in Eq.(\ref{eq:kappa_eq})
and in Eq.(\ref{eq:rho_eq}) to obtain the lowest order 
contributions to the pairing function and to the one-body density matrix:
\begin{equation}
\bar \kappa(\vec{r}_1,\vec{r}_2)\simeq \sqrt{n} \ \phi_0(r) \quad
, \quad 
\rho(\vec{r}_1,\vec{r}_2)\simeq n\exp\left(-\frac{r}{a}\right) \quad .
\label{eq:kappa0_ro0}
\end{equation}
This expression of $\bar \kappa(\vec{r}_1,\vec{r}_2)$ clearly shows that the
pairing function describe the spatial structure of two atoms linked in the
molecular bound state $\phi_0$. Similar results have been obtained in
\cite{Nozieres} for a different model potential.

	For a small but finite gaseous parameter $na^3$ the molecular 
condensate is not an ideal gas but rather a weakly interacting Bose
gas, with an effective coupling constant $g_{\rm mol}$ between the
molecules. We derive this coupling constant
by calculating the first correction to the expression Eq.(\ref{eq:brisure}) for the chemical potential.
We expand the expression Eq.(\ref{eq:spectrum}) to first order in $n a^3$ and substitute
the result in Eq.(\ref{eq:gap}). This leads to 
\begin{equation}
\mu = \frac{E_0}{2} (1-12\pi na^3+\ldots) \label{eq:mu} \quad .
\end{equation}
On the other hand the molecular chemical potential $\mu_{\rm mol}$, equal to
twice the atomic chemical potential $\mu$, is given in the usual mean-field approach
for condensates by 
\begin{equation}
\mu_{\rm mol}=E_0+g_{\rm mol} n_{\rm mol} \quad , \label{eq:mu_mol}
\end{equation}
where $n_{\rm mol} = n/2$ is the density of molecules. We
therefore deduce for the coupling constant between molecules \cite{simple}:
\begin{equation}
g_{\rm mol}= 6 g\quad . \label{eq:central}
\end{equation}
As $g_{\rm mol}$ is positive the molecular condensate is stable with respect to a
spatial collapse.

	The value of the molecular coupling constant $g_{\rm mol}$ can also
be obtained from the response of the gas to a time dependent perturbation. It
is important to check that the corresponding value coincides with the static
prediction Eq.(\ref{eq:central}).
Imagine that one perturbs the system from thermodynamical equilibrium 
by applying an external potential on the atoms for a finite time
interval $[0,\tau]$. We describe the evolution of the gas by a time
dependent Gaussian ansatz for the many-body density operator \cite{Blaizot}.
In that case, as in \cite{BCS}, we deduce from the Heisenberg equation for the field
operator and from  Wick's theorem the time evolution of the pairing function,
written here for convenience for $t>\tau$:
\begin{eqnarray}
\left[-i\hbar \partial_t\right. &-&\left. \frac{\hbar^2}{4m}\Delta_{\vec R} 
+2(gn(\vec{r}_1)+gn(\vec{r}_2)-\mu)
+{\cal H}_{r}\right] \bar \kappa(\vec{r}_1,\vec{r}_2) \nonumber \\
&=& -g \bar \kappa_{\rm reg}(\vec{r}_1)\rho^*(\vec{r}_1,\vec{r}_2)
-g \bar \kappa_{\rm reg}(\vec{r}_2)\rho(\vec{r}_1,\vec{r}_2) \quad . \label{eq:tdhfb}
\end{eqnarray}
We assume that the applied perturbation varies very slowly spatially
at the scale of the scattering length $a$.
Thus the pairing function has negligible components on the scattering states and
can be assumed to have the same $r$ dependence as $\phi_0$. We therefore take
the Local Density Approximation:
\begin{equation}
\bar \kappa(\vec{r}_1,\vec{r}_2) = \left[n(\vec{R},t) \right]^{1/2}
\phi_0(r) \ \exp\left[iS(\vec{R},t)\right] \quad .
\end{equation}
We close equation (\ref{eq:tdhfb}) with the Local Density
Approximation for the one-body density matrix: 
\begin{equation}
\rho(\vec{r}_1,\vec{r}_2) = n(\vec{R},t) \ \exp\left(-\frac{r}{a}\right)
\quad . \nonumber
\end{equation}
A simple projection of Eq.(\ref{eq:tdhfb}) on the bound
state (\ref{eq:bound}) leads to the Gross-Pitaevskii equation 
\begin{equation}
 i \hbar \partial_t \psi_P = \left( - \frac{\hbar^2}{4m} \Delta_{\vec{R}} + 6 g |\psi_P|^2 - 2 \mu + E_0 \right)
\psi_P \quad ,\label{eq:pitaev}
\end{equation}
where we have introduced the macroscopic wave function
$\psi_P(\vec{R},t) = \left(n/2\right)^{1/2} \ \exp(iS)$ describing the molecular 
condensate. We note that this equation, confirms in a direct way our
previous finding on $\mu$ at equilibrium (Eq.(\ref{eq:mu})). From the
linear analysis of Eq.(\ref{eq:pitaev}), we predict a Bogoliubov spectrum for the
molecular condensate different from the atomic condensate:
\begin{equation}
	\hbar \omega_P(k) = \left( \frac{\hbar^2 k^2}{4m} \right)^{1/2}
	\left(\frac{\hbar^2 k^2}{4m} + 6 g n \right)^{1/2} \quad .
\end{equation}
	
	Measurement of this spectrum could be used as an experimental
evidence for the condensation of pairs.

	As a conclusion let us stress three points. First, one word about the temperature. Indeed,
the critical temperature at which the condensate of molecules (mass
$2m$, density $n/2$) forms is 
\begin{equation}
	k_BT_c = \frac{\pi \hbar^2}{m}
\left[\frac{n}{2\zeta(3/2)}\right]^{2/3} \quad .
\end{equation} 
In the low density regime $k_BT_c \ll |E_0|$, so that condensation of
pairs can occur without any thermal dissociation. Second, it would be
interesting to test the prediction on the coupling constant  ($6g$)
by a direct calculation of the scattering of two molecules of the type
considered here. Finally, our model does not describe the relative 
stability of this molecular condensate toward the formation of
molecules in deeper bound levels. An evaluation of the creation rate
of deep bound states by collision of an atom with one molecule would 
be a relevant complement to this analysis.

	The author thanks Yvan Castin for helpful contribution to this
work. The author also wishes to thank D.~Vautherin for stimulating
discussions and the  members of the LPTL for their warm
welcome in the laboratory.


\begin{thebibliography}{99}

\bibitem{BEC}
M. H. Anderson, J.R. Ensher, M.R. Matthews, C.E.
Wieman, and E.A. Cornell, Science {\bf 269}, 198 (1995);
K.B. Davis, M.O. Mewes, M.R. Andrews, N.J. van Druten,
D.S. Durfee, D.M. Kurn, and W. Ketterle, 
Phys.\ Rev.\ Lett. {\bf 75}, 3969 (1995);
C.C. Bradley, C.A. Sackett, and R.G. Hulet, Phys. Rev. Lett. {\bf 78},
985 (1997).

\bibitem{Revue} F. Dafolvo, S. Giorgini, L.Pitaevskii, and S. Stringari,
Rev. Mod. Phys. {\bf 71}, 463 (1999).

\bibitem{Boston} R.~de~Carvalho, J.M.~Doyle, B.~Friedrich, T.~Guillet,
J.~Kim, D.~Patterson and J.D.~Weinstein, Eur. Phys D {\bf 7}, 289 (1999).

\bibitem{Pillet}  A. Fioretti, D. Comparat, A. Crubellier, O. Dulieu,
F. Masnou-Seeuws, and P. Pillet, Phys. Rev. Lett. {\bf 80} 4402 (1998).

\bibitem{Heinzen} R. Wynar, R.S.~Freeland, D.J.~Han, C.~Ryu, and D.J.~Heinzen
Science {\bf 287}, 1016 (2000). 

\bibitem{Landau} C.J. Joachain, {\sl Quantum Collision Theory}
(North-Holland, 1983), pp. 78$-$105; L.D. Landau and E.M. Lifshitz, 
{\it Quantum Mechanics} (Pergamon Press, 1977), {\S} 133.

\bibitem{Shlyap} P.O. Fedichev M.W. Reynolds and G.V. Shlyapnikov, Phys. Rev. Lett. {\bf 77}, 2921 (1996).

\bibitem{Greene} B.D. Esry, C.H. Greene and J.P. Burke Jr., Phys. Rev. Lett. {\bf 83}, 1751 (1999).

\bibitem{Ketterle_dipole} D.M.~Stamper-Kurn, M.R.~Andrews,
A.P.Chikkatur, S.~Inouye, H.J.~Miesner, J.~Stenger and W.~Ketterle, Phys. Rev. Lett. {\bf 80} 2027 (1998).

\bibitem{Ketterle_Feshbach} S.~Inouye, M.R.~Andrews, J.~Stenger,
H.-J.~Miesner, D.M.~Stamper-Kurn, W.~Ketterle, Nature {\bf 392} 151 (1998).

\bibitem{Cornell_Feshbach} S.L.~Cornish, N.R.~Claussen, J.L.~Roberts,
E.A.~Cornell, C.E.~Wieman, Preprint.

\bibitem{Huang} K.\ Huang, \textit{Statistical Mechanics} 
(Wiley, NY, 1987).

\bibitem{BCS} G. Bruun, Y. Castin, R. Dum and K. Burnett,
Euro. Phys. J. D {\bf 7}, 433 (1999).

\bibitem{Blaizot} J.P. Blaizot and G. Ripka {\sl Quantum Theory of
Finite Systems} (The MIT Press, 1985) chapter 7 for the equilibrium
case and chapter 9 for the time dependent case.

\bibitem{Trap} In a real trap, the trapping
potential $U(\vec{R}\,)$ should be added to Eq.(\ref{eq:h_eq},\ref{eq:Phi}).

\bibitem{warning} The eigenvalues $\hbar \omega_k$ are the ``best'' parameters for the 
Gaussian ansatz and should not be considered as energies of any collective mode. 

\bibitem{Nozieres} P. Nozi{\`e}res and D. Saint James, Journal de Physique
{\bf 43}, 1133 (1982).

\bibitem{gap} When the atomic condensate vanishes, $\Phi=0$ and
Eq.(\ref{eq:gap}) does not provide an explicit expression for $\bar \kappa_{\rm
reg}$.

\bibitem{simple} We may consider this problem in a completely
different way. Let us take indeed four indiscernables atoms  interacting
with the potential Eq.(\ref{eq:vpseudo}) and  plunged into a box
of volume $L^3$. We choose as a trial ground state wave function $
\langle \vec{r}_1, \vec{r}_2, \vec{r}_3, \vec{r}_4 |\psi_4 \rangle 
= {\mathcal N} {\mathcal S}\left[\phi_0( \vec{r}_1-\vec{r}_2)
\phi_0(\vec{r}_3-\vec{r}_4)\right]$, where ${\mathcal S}$ is the
operator of symmetrization and $ {\mathcal N}$ a
normalization constant. This state describes four atoms linked in
two pairs. Then, a simple calculation leads to the
mean energy $E = \langle \psi_4 | H | \psi_4  \rangle = 2 E_0 + \frac{6 g}{L^3}$.

\end{thebibliography}
\end{document}